\documentclass[aip,twocolumn,preprintnumbers,amsmath,amssymb]{revtex4}
\usepackage{graphicx}
\usepackage{dcolumn}
\usepackage{bm}
\usepackage{mhchem}
\usepackage{textcomp}

\usepackage{multirow}
\usepackage{longtable}
\usepackage{supertabular}
\usepackage{array}
\usepackage{booktabs}
\usepackage{subfigure}
\usepackage{lscape}

\begin{document}

\title{Accurate atomic quantum defects from particle-particle random phase approximation}

\author{Yang Yang}
\email{yy88@duke.edu}
\affiliation{Department of Chemistry, Duke University, Durham, NC 27708, U.S. }

\author{Kieron Burke}
\email{kieron@uci.edu}
\affiliation{Department of Chemistry, University of California, Irvine, CA 92697,  USA}

\author{Weitao Yang }
\email{weitao.yang@duke.edu}
\affiliation{Department of Chemistry, Duke University, Durham, NC 27708, U.S. }
\affiliation{Department of Physics, Duke University, Durham, NC 27708, U.S.}
\affiliation{Key Laboratory of Theoretical Chemistry of Environment, Ministry of Education,
School of Chemistry and Environment, South China Normal University, Guangzhou 510006, China}


\date{\today}

\def\ben{\begin{equation}}
\def\een{\end{equation}}
\begin{abstract}
The accuracy of calculations of atomic Rydberg excitations cannot be judged
by the usual measures, such as mean unsigned errors of many transitions.
We show how to use quantum defect theory to (a) separate errors due to
approximate ionization potentials, (b) extract smooth quantum defects to
compare with experiment, and (c) quantify those defects with a few 
characteristic parameters.
The particle-particle random phase approximation (pp-RPA) produces
excellent Rydberg transitions that are an order of magnitude more accurate
than
those of time-dependent density functional theory with standard
approximations.  We even extract reasonably accurate defects from the lithium
Rydberg series, despite the reference being open-shell.  Our methodology
can be applied to any Rydberg series of excitations with 4 transitions
or more to extract the underlying threshold energy and characteristic quantum
defect parameters.  Our pp-RPA results set a demanding challenge for other
excitation methods to match.

\end{abstract}

\maketitle

\section{\label{Introduction}Introduction}

An accurate description of electronic excited states has
always been a major goal for theoretical and computational chemists.
Nowadays, linear-response time-dependent density functional theory (LR-TDDFT)
\cite{Runge_1984_997,Casida1996,Petersilka_1996_1212,Ullrich2012} has
become the standard workhorse for practical applications,
because of the favorable balance between
accuracy and computational cost. 
However, there are a variety of well-established limitations of TDDFT
with the standard semilocal functionals \cite{Burke_2005_62206}. 
These challenges include double excitations, charge transfer (CT)
excitations, and Rydberg excitations\cite{Tozer_2000_2117, Cave_2004_39,
Tozer_1999_859, Dreuw_2003_2943, Dreuw_2005_4009, Ullrich2012}. 
Although many of these problems can be ameliorated by
using range-separated \cite{Leininger_1997_151, Iikura_2001_3540} or more accurate asymptotic-corrected
response kernels \cite{Leeuwen_1994_2421,Tozer_1998_10180,Casida_2000_8918,Wu_2003_2978}, the results are not always 
satisfactory,
and a more careful examination beyond the
usual tabulations of mean absolute errors is required to check
for an accurate Rydberg series.

Quantum defect (QD) theory\cite{Seaton_1983_167} has long been applied
as a test for atomic Rydberg excitations\cite{Faassen_2006_94102}.
In the long history of quantum mechanics, the QD
was originally introduced to empirically describe 
weak electron penetration effects relative to
strong shielding effect \cite{Sommerfeld_1916_1, Seaton_1983_167}. Later, systematic derivations were
given, creating a non-empirical theory \cite{Seaton_1983_167}. 
The QD can also be applied to describe many molecular
Rydberg states \cite{Guerout_2004_3043, Guerout_2004_3057, Petsalakis_2003_2004, Merkt_1997_675, Herzberg_1987_27}. 

In the present work, we restrict ourselves to atomic Rydberg series.
QDs are often given as lists, with the index corresponding
to the principal quantum number.  But the crucial feature of 
QDs is that they are smoothly varying functions of the energy.
In fact, they even merge continuously with scattering phase shifts
across the ionization threshold\cite{FB09}. We use the
smoothness feature to create our procedure for extracting
QDs from lists of excitation energies.  In earlier work,
Van Faassen and Burke \cite{Faassen_2006_94102} analyzed the singlet and
triplet spin-states of the S- and P- series for atomic helium, 
beryllium and neon as test cases for standard TDDFT. 
Although excitation energy errors were no more than
a few millihartree, the delicate QD results exposed
their limitations.   With a ground-state DFT calculation that yields
a potential with the correct asymptotic decay (although see Ref. \cite{WBb05} for
a way around even this restriction), QDs that were reasonably
but not highly accurate were usually found with semilocal approximations
to the kernel in TDDFT. Later, the D series in Be was shown to be
very poorly described by TDDFT\cite{Faassen_2006_410}.

Recently, two of us developed the particle-particle random
phase approximation (pp-RPA) and the particle-particle Tamm-Dancoff
approximation (pp-TDA) theory for calculation of challenging excitations such
as double excitations, charge transfer excitations, Rydberg excitations,
and excitations in diradicals. \cite{Yang_2013_224105, Yang_2014_124104,
Yang_2015_4923}.  
Good excitation energies were typically found with a relatively
low $O(N^4)$ computational cost, where $N$
is the number of virtual orbitals.\cite{Yang_2014_124104}
Thus pp-RPA and pp-TDA are promising methods that complement standard LR-TDDFT,
especially for these challenging cases.  
In the present work, we develop QD analysis to
more stringently test these methods, and compare them to TDDFT.

\begin{table}
\caption[]{\label{BeExci} First 11 excitation energies in Be $^1$P series.
For transitions to $n=6$ and higher, the values have been extracted
from the fits to the associated QDs given later (for ALDA, this begins
at $n=5$).}
\begin{ruledtabular}
\begin{tabular}{cccc}
Transition & Expt. & pp-TDA & ALDA\tabularnewline
\hline
2s$\rightarrow$2p & 0.1940  & 0.1959  & 0.1868\tabularnewline
2s$\rightarrow$3p & 0.2742  & 0.2737  & 0.2710\tabularnewline
2s$\rightarrow$4p & 0.3054  & 0.3047  & 0.3048\tabularnewline
2s$\rightarrow$5p & 0.3195  & 0.3186  & 0.3194  \tabularnewline
\hline
MSE (mH) &  & -0.1 & -2.8\tabularnewline
MUE (mH) &  & 1.1 & 2.8\tabularnewline
\hline
2s$\rightarrow$6p & 0.3269  & 0.3259  & 0.3269 \tabularnewline
2s$\rightarrow$7p & 0.3313  & 0.3302  & 0.3313 \tabularnewline
2s$\rightarrow$8p & 0.3340  & 0.3329  & 0.3341 \tabularnewline
2s$\rightarrow$9p & 0.3359  & 0.3348  & 0.3359 \tabularnewline
2s$\rightarrow$10p & 0.3372  & 0.3361  & 0.3372 \tabularnewline
2s$\rightarrow$11p & 0.3382  &  0.3370 & 0.3382\tabularnewline
2s$\rightarrow$12p &  0.3389 & 0.3378  & 0.3389\tabularnewline
\hline
MSE (mH) &  & -1.1 & 0.0\tabularnewline
MUE (mH) &  & 1.1 & 0.0\tabularnewline
\hline
All \tabularnewline
MSE (mH) &  & -0.7 & -1.0\tabularnewline
MUE (mH) &  & 1.1 & 1.0\tabularnewline
\end{tabular}
\end{ruledtabular}
\end{table}
To illustrate the need for the QD analysis, we first
analyze a Rydberg series using the traditional methods of
measuring errors in electronic structure methods.  In Table \ref{BeExci}, we
list the lowest 11 excitation energies in the Be P singlet series.
We include the experimental values, the pp-TDA results (details given
later) and TDDFT calculations using the ALDA kernel and
{\em exact} ground-state KS potential (see Ref. \cite{Faassen_2006_94102}).   
We see that, for the lowest transition frequencies, pp-TDA is far better
than ALDA.  But beyond about the fifth transition, the pp-TDA error
stops decreasing, while the ALDA error keeps getting smaller.
Thus, when we average over transitions 2-5, pp-TDA is clearly much
better than ALDA.  But when average over 6-12, the order has been
reversed, so that ALDA now appears better when averaged over all
transition.  This trend would continue, with the mean errors in ALDA
going to zero as the total number of transitions included increases,
while that of pp-TDA tends to about 1 mH.  
However, we show below that this is entirely an
artifact of the error in the ionization potential (IP) of pp-TDA, which is absent in the TDDFT
calculations by virtue of using the exact KS potential.  As the number
of transitions included increases, the errors reflect simply the error
in the IP.  For any finite number, there is no way to separate the two
effects by this means.  In fact, we show below that the pp-TDA results listed here
are almost an order of magnitude better than the TDDFT results.

\section{\label{Theory}Theory}

\subsection{\label{TheoryQD} Quantum defect theory as a measure of Rydberg excitations}

Consider a Rydberg series of excited states of a neutral atom, with energies
$E_{l,S}(n)$ below the ionization threshold, 
where $n$ runs from the first allowed excitation and is unbounded.
The QD for this series is defined by:
\begin{equation}
\label{QDDef1}
E_{l,S}(n) 
=-\frac{1}{2(n-\mu_{l,S}(n))^2},
\end{equation}
i.e., as $n$ grows, $E$ approaches 0 ever more slowly, mimicking the behavior
of a H-atom series, but with small deviations.  The QD is a dimensionless
measure of the deviation from a pure Hydrogenic series. 

The key ingredient of our analysis is to generalize 
the QD as a function of index $n$ to
a continuous function of $E$, i.e., $\mu(E)$ \cite{Faassen_2006_94102, Faassen_2006_410}, with the
requirement that 
\ben
\mu(n)=\mu(E(n)).
\een
In fact, all QDs are smooth functions of $E$ in practice.  We make
the further assumption that, on the scale of energies spanned by the series, $\mu$
is not strongly varying, and can usually be well-approximated by a simple
parabola:
\ben
\label{mufit}
\mu (E) \approx a  + \,b E + c\,  E^2.
\een
Thus an entire, infinite Rydberg series can be very accurately represented
by three real numbers, one of which ($a$) is simply the quantum defect at
threshold $\mu_0$, and the accuracy of an approximate Rydberg series can be judged
by the accuracy of its approximation for $a,b,c$.  For the rest of this paper,
we approximate all such curves by parabolas.

But transition frequencies are measured or calculated relative to the
ground state, so a fourth number, the IP, enters:
\ben
\omega_n = I - \frac{1}{2(n-\mu_n)^2},
\een
or
\ben
\label{QDDef2}
\mu_n = n - \frac{1}{\sqrt{2(I-\omega_n)}}.
\een
While $I$ might be known very accurately for a given experiment, in 
calculations its precise value is slightly affected by the limitations
of a calculation, such as finite basis sets.  As $n$ grows, even tiny 
errors in $I$ will cause $\mu_n$ to become highly inaccurate.  Thus our
procedure is designed to accommodate uncertainty in the value of $I$, and
below we test this in cases where $I$ is known.

Expanding $\mu(E)$ in a Taylor series around $E=0$, we find
\ben
\label{QDfitapprox}
\omega_n = I -[2(x-(\tilde b-\tilde c/(2x^2)/(2x^2)]^{-1},~~~~x=n-\tilde a.
\een
Thus, given at least 4 transition frequencies, we can fit a Rydberg
series to find the best four parameters, which gives us our
best estimate for $I$.    Note that we have added a tilde to
each of $a,b,c$, because these parameters are {\em not} best
estimated by this procedure, as we performed an expansion around $E=0$
to find Eq. (\ref{QDfitapprox}).
Thus, in a second fitting step, we fix $I$, and refit $a,b,c$ via
Eqs. (\ref{mufit}-\ref{QDDef2}), yielding our best estimate for these.
Once we have values for $a,b,c$, we plot
the QD as a function of $E$ and see how well a method performs.

\begin{figure}[ht]
 \centering
	  \includegraphics[width=0.5\textwidth]{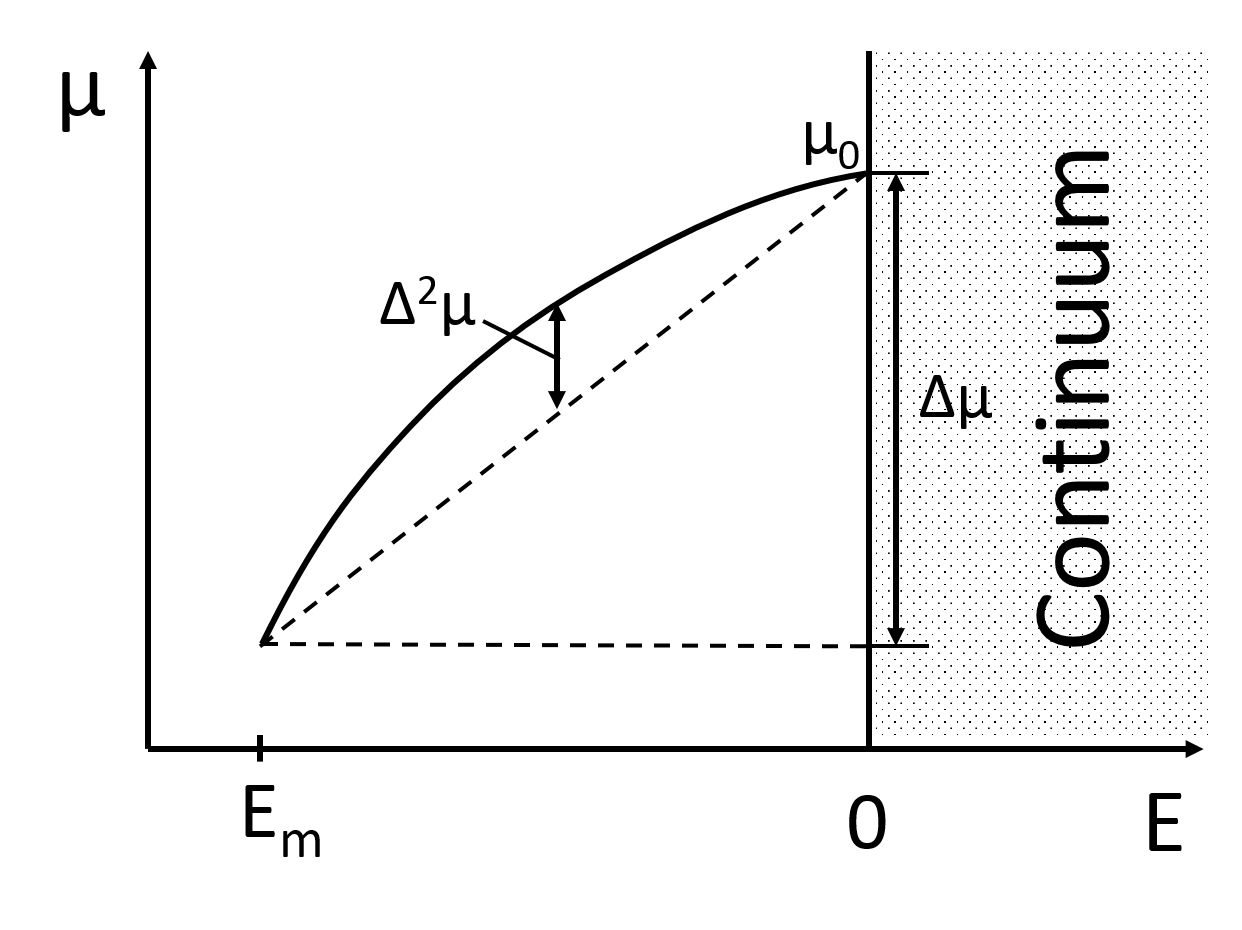}
\caption[]{\label{fig:illustration} Geometric illustration of the physical meanings of QD parameters.}
\end{figure}

\def\dmu{\Delta\mu}
\def\d2mu{\Delta^2\mu}
A final piece of methodology is to convert the parameters $a,b,c$ to
others that are easier to interpret.   Denote by $E_m$ the minimum value of
$E$, i.e, the lowest transition in the series.  We then write
\ben
\mu(E) = \mu_0 + x\, \dmu\,  + 4\, x(1-x)\, \d2mu,~~~~x=E/E_m.
\een
The parameters have been carefully chosen to have an immediate
physical interpretation.  Here 
$\mu_0$ is the QD at threshold, while $\dmu$,
the QD shift, is the
change in QD from the lowest transition to the threshold.
A negative value means the QD drops, a positive value means it is
rising, and $\mu_0+\dmu$ is the fit value of $\mu(E_m)$.
Lastly,   we call
$\d2mu$ the QD curvature, and is  the maximum
deviation from linearity, which occurs at $E_m/2$.
A negative value means the curve is convex, a positive value
means concave.
This simple geometric interpretation is shown in Fig \ref{fig:illustration}.
We will use these
values to judge and interpret the accuracy of approximate calculations
of Rydberg series.

\subsection{\label{TheoryppRPA} Basic theory on pp-RPA}

The pp-RPA can be derived in a variety of independent ways,
including via the adiabatic connection-pairing matrix 
fluctuations\cite{Aggelen_2013_30501, Aggelen_2014_18A511}, 
equations of motion \cite{Rowe_1968_153, Ring2004, Yang_2013_174110},
and time-dependent density functional theory with a
pairing field \cite{Peng_2014_18A522}. The key working equation
of pp-RPA is a generalized eigenvalue equation
\begin{equation}
\label{matrixequation}
\begin{aligned}
\begin{bmatrix}
\mathbf{A} & \mathbf{B}\\
\mathbf{B}^\dagger & \mathbf{C}
\end{bmatrix}
\begin{bmatrix}
\mathbf{X}\\
\mathbf{Y}
\end{bmatrix}
=\omega^{N\pm2}
\begin{bmatrix}
\mathbf{I} & \mathbf{0}\\
\mathbf{0} & -\mathbf{I}
\end{bmatrix}
\begin{bmatrix}
\mathbf{X}\\
\mathbf{Y}
\end{bmatrix},
\end{aligned}
\end{equation}
with
\begin{equation}
\label{eq:ppRPAdetail}
\begin{aligned}
A_{ab,cd}=&\delta_{ac}\delta_{bd}(\epsilon_a+\epsilon_b)+\langle ab||cd\rangle\\
B_{ab,kl}=&\langle ab||kl\rangle\\
C_{ij,kl}=&-\delta_{ik}\delta_{jl}(\epsilon_i+\epsilon_j)+\langle ij||kl\rangle,
\end{aligned}
\end{equation}
where $a,b,c,d$ are virtual orbital
indices and $i,j,k,l$ are occupied orbital indices
with the restrictions that $a>b$, $c>d$, $i>j$ and $k>l$. The brackets are defined as
\begin{equation}
\begin{aligned}
\langle p q|r s\rangle\equiv\int d\mathbf{r}_1d\mathbf{r}_2
\frac{\phi_p^*(\mathbf{r}_1)\phi_q^*(\mathbf{r}_2)
\phi_r(\mathbf{r}_1)\phi_s(\mathbf{r}_2)}
{|\mathbf{r}_1-\mathbf{r}_2|},
\end{aligned}
\end{equation}
and $\langle p q||r s\rangle\equiv\langle p q|r s\rangle-\langle p q|s r\rangle$.
This equation describes the transition process of adding or removing 
{\em two} electrons from a system. 
In an excitation energy calculation, we usually adopt a 
two-electron deficient reference and investigate its two-electron
addition processes,  yielding a series of neutral states including
both the ground and electronically excited states. 
Differences between transition energies are
neutral excitation energies with 
\begin{equation}
\label{dataprocess}
E_{0\rightarrow n}^N=(E_n^N-E_0^{N-2})-(E_0^N-E_0^{N-2}),
\end{equation}
where $N$ and $N-2$ denote the number of electrons, and
$0$ and $n$ denote the ground and excited states, respectively. 
By solving Eq. \eqref{matrixequation} and processing 
the output with Eq. \eqref{dataprocess}, we obtain 
the excitation energies for the $N$-electron system.

There is a corresponding Tamm-Dancoff approximation (pp-TDA)
to pp-RPA.  We simply set $\mathbf{B}=\mathbf{C}=0$.
The result of pp-TDA is often similar to pp-RPA, and the difference is negligible in small systems with limited number of electrons. Therefore, we use the pp-TDA throughout this paper, which is slightly cheaper than pp-RPA.

\section{\label{Computational details}Computational details}

We choose both singlet (S=0) and triplet (S=1) Rydberg series for
all our tests, so we are probing both spin-conserving and spin-flipping
excitations with respect to the ground state.   We look at three different angular momenta: S, P, D,
to further test the pp-RPA method.  Lastly, we consider two different
closed-shell atoms, Be and Mg, whose QDs are very distinct,
and also apply our methods to Li, to see the challenges of an open-shell
ground state.
The difference between pp-TDA and pp-RPA is almost undetectable for 
these species, so we report data computed with pp-TDA. 
We use HF for the reference.
A very
extensive even-tempered basis set was built\cite{Helgaker2000,Wu_2005_711}
with exponents satisfying $\alpha_i=\alpha_1\beta^{i-1}$. Each basis
contains 22s, 18p and 17d functions with the smallest exponents being
$\alpha_1=$ 0.0002441406, 0.0004882813, 0.0002441406, respectively. 
The basis has been tested to give highly converged excitation energies
($\Delta E < 0.1$ mH) for diffuse Rydberg states as high as $n=6$.
All calculations are performed with Cartesian basis set on
the QM4D package \cite{QM4D}. 

\section{\label{Results}Results}

In this section, we systematically dissect the results of our
calculations.

\subsection{\label{validity} Extracting ionization potentials}

Since the IP of a given, calculated Rydberg series
is often not reported (or possibly not even calculated), in this section
we demonstrate that our fitting procedure provides a method for
{\em extracting} accurate IPs from such a series.

\begin{table}
\caption[]{\label{IPfitResult1} Errors (in mH) in fitted ionization potential for Be and Mg
Rydberg series; {\em Exp} is experimental series, {\em pp} is the pp-TDA calculation with
HF reference, and {\em avg} is the average of all 6 estimated IP's. }
\begin{ruledtabular}
\begin{tabular}{|c|c|ccc|ccc|c|}
 &  & $^1$S & $^1$P & $^1$D & $^3$S & $^3$P & $^3$D & avg \tabularnewline
\hline
\multirow{2}{*}{Exp} & Be & 0.011 & -0.008 & -0.006 & 0.024 & 0.073 & 0.032 & 0.021\tabularnewline
 & Mg & -0.005 & -0.003 & -0.034 & 0.025 & 0.052 & 0.017 & 0.009\tabularnewline
\hline
\multirow{2}{*}{pp}
 & Be & -1.057  & -1.173  & -0.958  & -1.053  & -1.002  & -1.038  & -1.047 \tabularnewline
 & Mg & -5.172  & -5.148  & -5.130  & -5.174  & -5.102  & -5.161  & -5.148 \tabularnewline
\end{tabular}
\end{ruledtabular}
\end{table}
We do this by applying our procedure to the {\em experimental}
transition frequencies, and comparing the fitted IP with the 
experimental value.   
The IPs of Be and Mg are 0.342603 and 0.280994 Hartree, respectively \cite{Kramida2012}.
For both Be and Mg, we have 6 separate Rydberg
series, and for Li we have three.  In Table \ref{IPfitResult1}, we list all the
errors in the IPs extracted from the fitting procedure.
In the rows marked Exp, we have fitted the IP in each Rydberg series,
and measure the error relative to the exact value.  
We see that errors are of order 0.05 mH or less, showing how accurately
the IP can be found from a Rydberg series.

The error in the averaged IP's from the series is
only about $5\times10^{-4}$ eV. These results not only confirm the validity of
our fit, but also show the magic of QD theory. 
With only four accurate data points in a Rydberg series, 
QD theory yields an IP of this
accuracy. Furthermore, in principle, if we plug the acquired parameters ($a$, $b$, $c$,
and IP) to Eq. \eqref{QDfitapprox}, we can predict
all the excitation energies in that Rydberg series.

In the second set of IP's, we see the values from the pp calculation.  Now
we can see that the average IP differs noticeably from the experimental
value (-1 mH for Be, -5 mH for Mg).  This reflects the error in the IP of the
underlying reference HF calculation.  But given this built-in error,
we see that the pp Rydberg series are extremely consistent.  None differ
by more than 0.1mH from the average value.  Thus, QD analysis is extracting
the underlying IP to within 0.1mH, i.e., it shows that the IP for the Be
reference calculation is 0.3416 Hartree, and for Mg it is 0.2759 Hartree.

\begin{table}
\caption[]{\label{IPfitResult2} Same as Table \ref{IPfitResult1} but
for Li doublet series.  Errors in IP in mH.}
\begin{ruledtabular}
\begin{tabular}{|c|ccc|c|}
& S & P & D & avg\tabularnewline
\hline 
Expt & 0.010  & 0.010 & 0.007 & 0.009\tabularnewline
pp-AB & 4.203  & 4.237  & 4.230  & 4.223 \tabularnewline
pp-BB & -3.252  & -3.219  & -3.225  & -3.232 \tabularnewline
pp-AVE & 0.476  & 0.509  & 0.502  & 0.496 \tabularnewline
\end{tabular}
\end{ruledtabular}
\end{table}
In Table \ref{IPfitResult2}, we give the results for 
the extremely challenging case of Li.  
The exact IP is 0.198142 Hartree \cite{Kramida2012}.
Now we can see
that each of the calculated series AB and BB (details given
later) has its own
distinct ionization threshold, each with noticeable errors.
However, when we average over transition frequencies,
the new series (AVE) has a much more accurate ionization threshold,
suggesting that is a good way to deal with the open-shell issue.
In fact, every transition energy of the AVE series is better
than the corresponding energy of either AB or BB.

\subsection{\label{sec:fitting} QD errors due to fitting}

In this section, we extract the QD parameters for the 
experimental Rydberg series, using the known IPs,
and also consider the calculated parameters
when the IP is fitted.  This gives us
a measure of the error introduced into the calculated QD's
due to that fitting.

In Table \ref{QDexpt}, we give the parameters for all the different Rydberg
series examined in this paper, using the experimental data.
The first point to notice is the value of the QD
at threshold.  These vary from about 1.6 down to about -0.1.  
These dimensionless numbers are very important, as knowing only
this value often yields a reasonably accurate Rydberg series,
and also determines the scattering cross-section in the low-energy
limit\cite{FWEZ07}.  It is clear that our series cover a large range
of different values for the QD.
\begin{table}
\caption[]{\label{QDexpt}QD parameters (times 100) obtained from experimental Rydberg 
series, both with and without experimental IP value.}
\begin{ruledtabular}
\begin{tabular}{|c|ccc|ccc|}
 \multirow{2}{*}{tr}& \multicolumn{3}{c|}{Expt IP} & \multicolumn{3}{c|}{Fit IP}\tabularnewline
 \cline{2-4}\cline{5-7}
& $\mu_0$ & $\Delta\mu$ & $\Delta^2\mu$ & $\mu_0$ & $\Delta\mu$ & $\Delta^2\mu$\tabularnewline
\hline 
\multicolumn{7}{|c|}{Be}\tabularnewline
\hline
1S & 67.1  & -1.6  & 0.1  & 67.4  & -1.3  & 0.2 \tabularnewline
3S & 77.2  & -4.9  & -0.0  & 77.8  & -4.3  & 0.2 \tabularnewline
1P & 36.7  & 20.1  & -2.1  & 36.6  & 20.0  & -2.1 \tabularnewline
3P & 36.0  & -20.4  & 2.2  & 37.0  & -19.4  & 2.9 \tabularnewline
1D & -10.2  & 9.2  & -1.9  & -10.4  & 8.9  & -2.0 \tabularnewline
3D & 10.4  & -0.6  & -0.2  & 11.6  & 0.6  & 0.2 \tabularnewline
\hline
MSE &  &  &  & 0.4  & 0.4  & 0.2 \tabularnewline
MUE &  &  &  & 0.6  & 0.5  & 0.3 \tabularnewline
\hline 
\multicolumn{7}{|c|}{Mg}\tabularnewline
\hline 
1S & 152.0  & -2.2  & 0.1  & 152.2  & -2.0  & 0.2 \tabularnewline
3S & 162.4  & -6.1  & 0.0  & 162.6  & -5.8  & 0.1 \tabularnewline
1P & 104.6  & 7.6  & -0.5  & 104.7  & 7.8  & -0.5 \tabularnewline
3P & 112.5  & -21.5  & 1.9  & 112.7  & -21.3  & 2.0 \tabularnewline
1D & 60.3  & 28.4  & 3.1  & 60.6  & 28.7  & 3.2 \tabularnewline
3D & 16.6  & -0.5  & 0.0  & 16.9  & -0.2  & 0.1 \tabularnewline
\hline
MSE &  &  &  & 0.2  & 0.2  & 0.1 \tabularnewline
MUE &  &  &  & 0.2  & 0.2  & 0.1 \tabularnewline
\hline 
\multicolumn{7}{|c|}{Li}\tabularnewline
\hline 
2S & 39.6  & -0.7  & -0.1  & 39.9  & -0.5  & -0.0 \tabularnewline
2P & 4.5  & 0.5  & -0.1  & 4.7  & 0.6  & -0.0 \tabularnewline
2D & -0.2  & -0.3  & -0.1  & 0.1  & 0.0  & -0.0 \tabularnewline
\hline
MSE &  &  &  & 0.3  & 0.3  & 0.1 \tabularnewline
MUE &  &  &  & 0.3  & 0.3  & 0.1 \tabularnewline
\end{tabular}
\end{ruledtabular}
\end{table}

Next, consider the values of $\dmu$.  These can have either sign,
meaning the QD can either increase or decrease with energy.  The
magnitude varies from almost 0 up to about 0.3.  This gives the
scale of the total change in the QD from the first transition to
the threshold.
Finally, we consider $\d2mu$.  These values are much smaller, 
never being larger than about 0.03, but also vary in sign.  Thus
this measure shows that the curvature has relatively
little effect on the actual value of QD, and therefore on the Rydberg
transition frequencies.

Now, we consider the values of the parameters that we found when the
threshold energy was fitted rather than known from experiment.
Typical errors introduced by the fitting procedure are
about 0.004 or less.  The only real outlier is the triplet D series in
Be, where the threshold value is in error by 0.01.  The curvature errors
are typically even smaller (0.001), but since their values are already
small, this appears as a greater fractional error.   We conclude
that errors on this scale will be introduced to QD parameters whenever
IP's are fit, as is done for all our pp calculations in the rest of
this paper.  Such errors are smaller than those made by the approximations
we study below.

\subsection{\label{sec:Be}Results for Be}

\begin{figure}[ht]
 \centering
	  \includegraphics[width=0.5\textwidth]{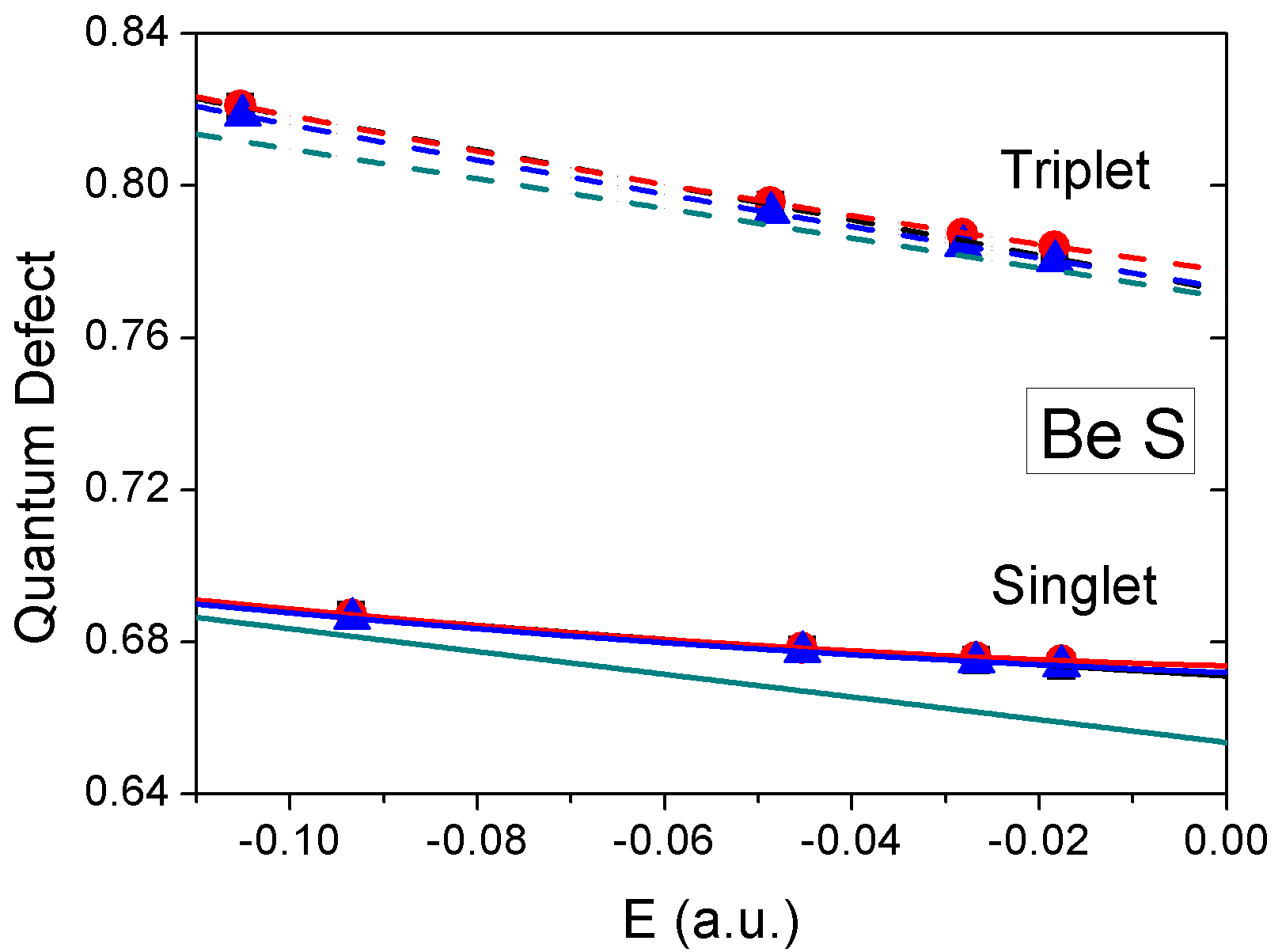}
\caption[]{\label{fig:BeS} S quantum defects for atomic Be  from experimental data (black),
with fitted IP (red), from pp-TDA (blue), and from TDLDA (green).}
\end{figure}
We begin the presentation of results with the simplest case:
singlet and triplet S excitations in Be.  We plot the experimental
and pp-TDA results in Fig. \ref{fig:BeS}, as well as the fit of
TDLDA results on the exact ground-state KS potential from Ref.
\cite{Faassen_2006_94102}.  The figure shows several
basic features.  First, the difference between the black lines and red
lines are almost invisible on this scale, showing that fitting the 
IP causes little degradation of results.   Second, the
pp blue lines are excellent approximations to the
red lines, indicating that pp-TDA produces near perfect QDs
for these series.  
Third, the TDDFT results (green) are
quite good, especially for the triplet.  

To quantify the differences in these curves, we extract 
$\mu_0$, $\dmu$ and $\d2mu$ from each curve in 
Tables \ref{QDexpt} and \ref{QDBeApp}, and compare the
calculations with the experiment (with the exact IP).
For both the triplet and singlet, the pp-TDA
values
are all excellent, and that even the curvature is within the error
of the fit.  The TDLDA results are almost as good (they were 
fit linearly\cite{Faassen_2006_94102}, because the curvature is so small).  However, there
is a significant error in the TDLDA threshold value for the singlet, as
is clearly visible in the figure, and the TDLDA slope is too large.

\begin{figure}[ht]
 \centering
	  \includegraphics[width=0.5\textwidth]{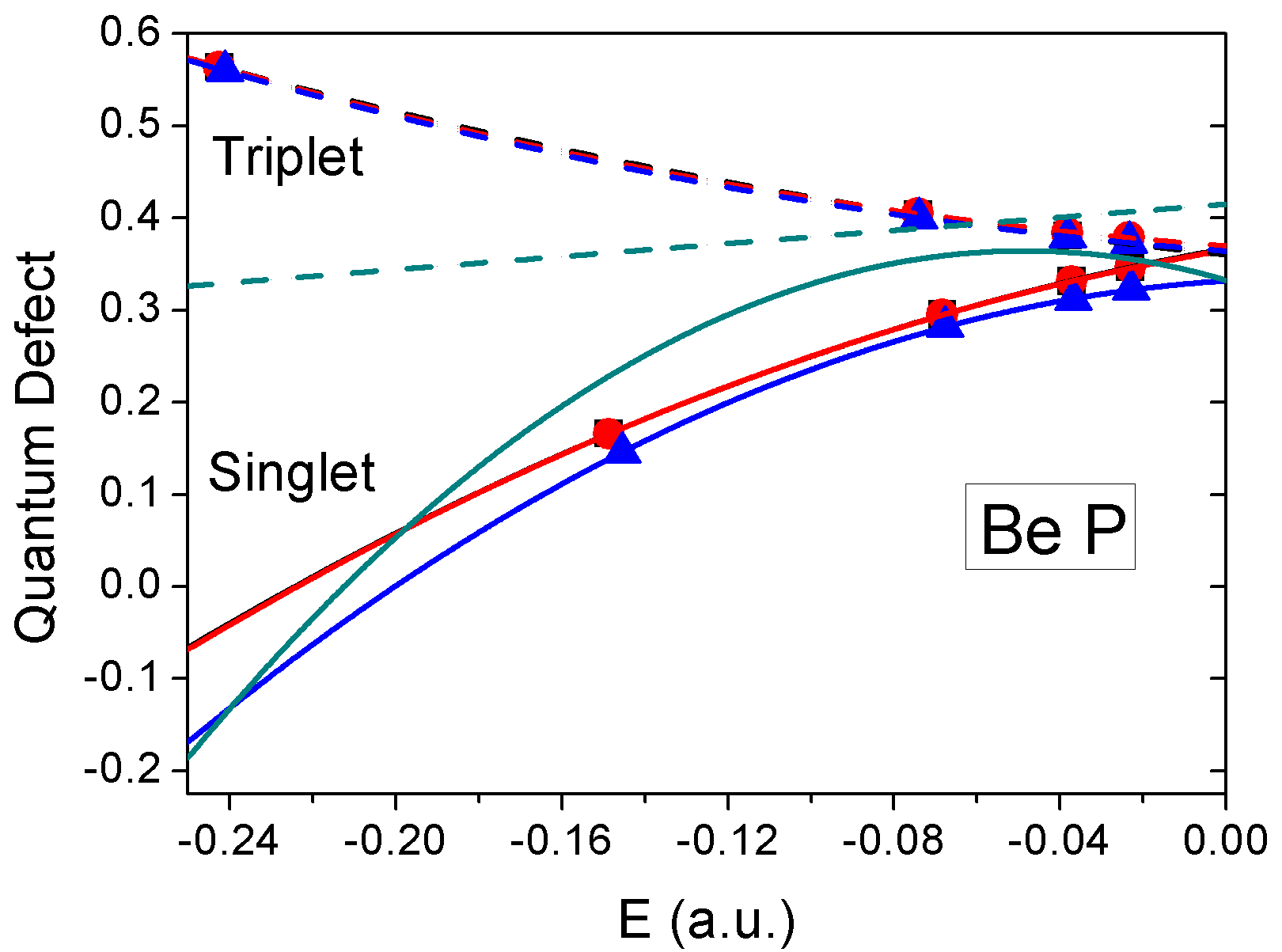}
\caption[]{\label{fig:BeP} Same as Fig. \ref{fig:BeS}, but for $P$.}
\end{figure}
\begin{figure}[ht]
 \centering
	  \includegraphics[width=0.5\textwidth]{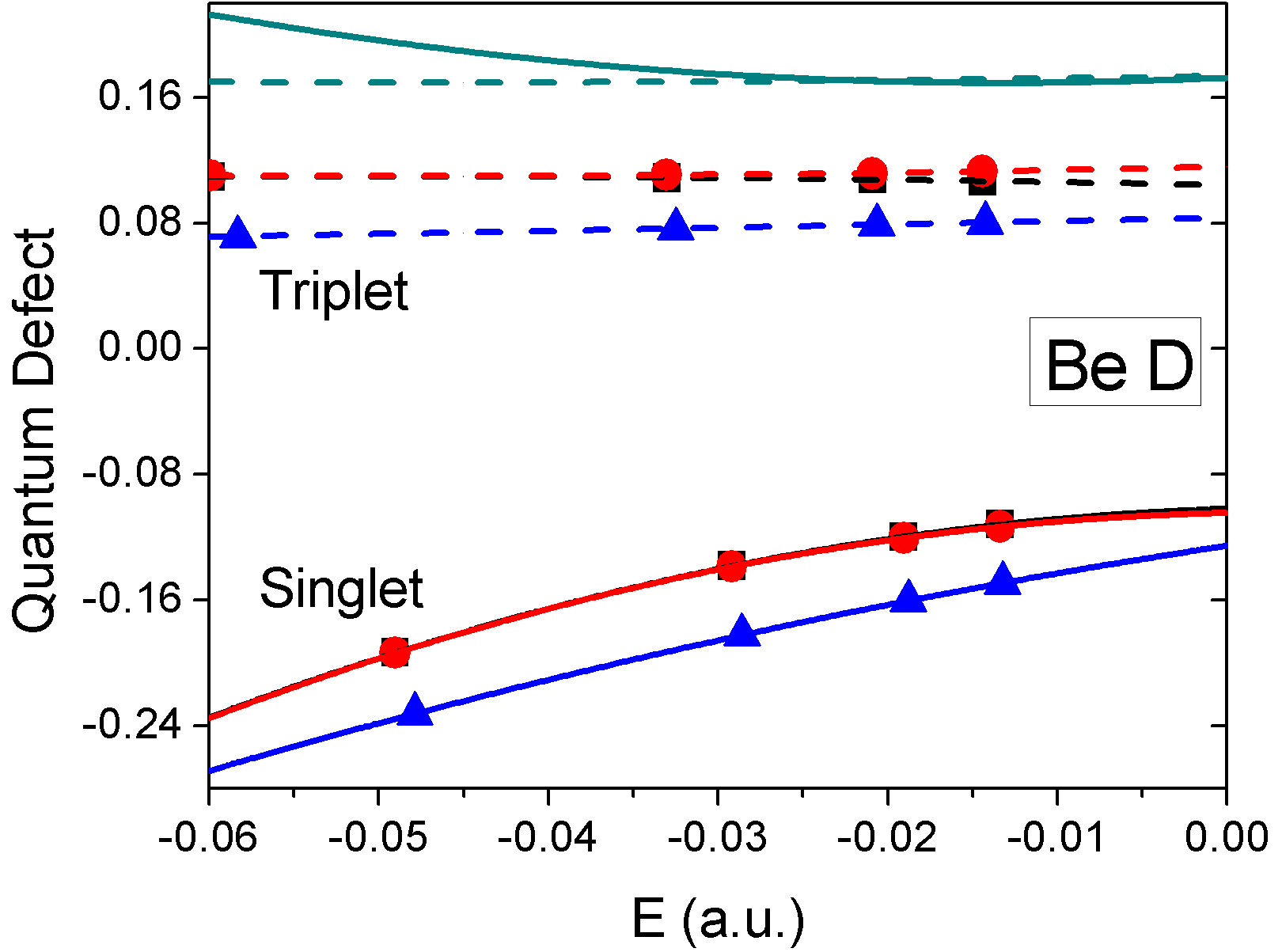}
\caption[]{\label{fig:BeD} Same as Fig. \ref{fig:BeS}, but for $D$.}
\end{figure}
So, does pp-RPA always produce such excellent Rydberg series?  
In the next two figures, we repeat the procedure for the P and D
series.  In the P case, we see that the experimental QDs
have substantially greater curvature than in the S series, especially
for the singlet, and that pp-RPA captures this effect well.  
On the other hand, TDLDA is not working nearly as well, with
the sign of $\dmu$ being incorrect for the triplet, and the
curvature being overestimated for the singlet.

The D QD has previously been noted as being
challenging\cite{Faassen_2006_410}, and here TDLDA fails entirely.  The TDLDA
singlet/triplet ordering is wrong, and the threshold value is 
incorrect by more than 0.2.  We see that although the 
pp-RPA results are less accurate than for S or P, they are
still qualitatively and even quantitatively correct. The
worst error is the threshold value for the triplet, being 
incorrect by 0.02, again, an order of magnitude better than
TDLDA. For the Be D series, pp-RPA succeeds where TDLDA fails.

\begin{table}
\caption[]{\label{QDBeApp} Same as Table \ref{QDexpt}, but for
approximate Be series.}
\begin{ruledtabular}
\begin{tabular}{|c|ccc|ccc|}
 \multirow{2}{*}{tr.} & \multicolumn{3}{c|}{pp} & \multicolumn{3}{c|}{ALDA}\tabularnewline
 \cline{2-4}\cline{5-7}
 & $\mu_0$ & $\Delta\mu$ & $\Delta^2\mu$ & $\mu_0$ & $\Delta\mu$ & $\Delta^2\mu$\tabularnewline
\hline 
1S & 67.2  & -1.4  & 0.1  & 65.4  & -2.8  & 0.0 \tabularnewline
3S & 77.3  & -4.5  & 0.1  & 77.1  & -4.1  & 0.0 \tabularnewline
1P & 33.2  & 18.6  & -3.7  & 33.3  & 12.4  & -8.2 \tabularnewline
3P & 36.4  & -19.6  & 2.8  & 41.5  & 9.0  & 0.0 \tabularnewline
1D & -12.5  & 10.7  & -0.7  & 17.3  & -5.0  & 2.1 \tabularnewline
3D & 8.3  & 1.2  & 0.00  & 17.4  & 0.4  & 0.2 \tabularnewline
\hline 
MSE & -1.2  & 0.5  & 0.1  & 5.8  & 1.3  & -0.7 \tabularnewline
MUE & 1.4  & 1.0  & 0.6  & 7.5  & 9.0  & 2.1 \tabularnewline
\end{tabular}
\end{ruledtabular}
\end{table}

Note that if a standard GGA or hybrid is used for
the ground-state calculation, TDLDA will not produce a Rydberg
series.  Even methods designed to enforce the correct asymptotic
behavior of the KS potential, such as the van Leeuwen-Baerends 
functional or asymptotically corrected methods, will typically
produce terrible QDs.  On the other hand, methods that include
exact exchange (often denoted EXX) produce excellent KS potentials,
whose performance will be comparable to that given here, since
the small error in IP does not destroy the QD results when properly
extracted.  But none will do better than the results here, since
we use the exact KS potential.

So, we can conclude that, for essentially every Be Rydberg
series, pp-RPA outperforms TDLDA, even when TDLDA has been applied
to the exact ground-state KS potential (and so the underlying
IP is exactly right).  
While ALDA does remarkably well for such a low-cost calculation,
it cannot compete with pp-RPA for accuracy.

\subsection{\label{Mg} Results for Mg}

\begin{figure}[ht]
 \centering
	  \includegraphics[width=0.5\textwidth]{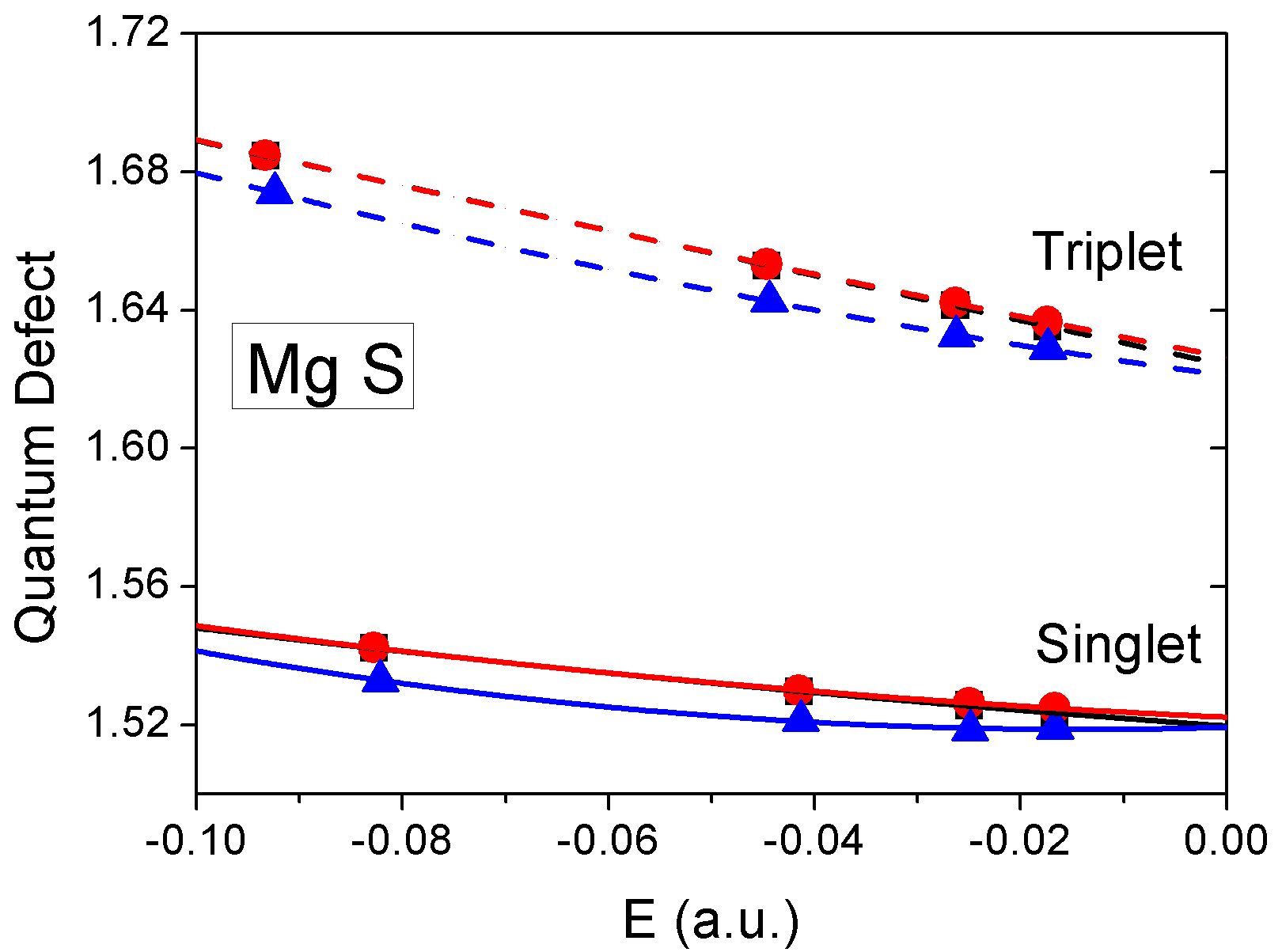}
\caption[]{\label{fig:MgS} Same as Fig. \ref{fig:BeS}, but for Mg.}
\end{figure}
\begin{figure}[ht]
 \centering
	  \includegraphics[width=0.5\textwidth]{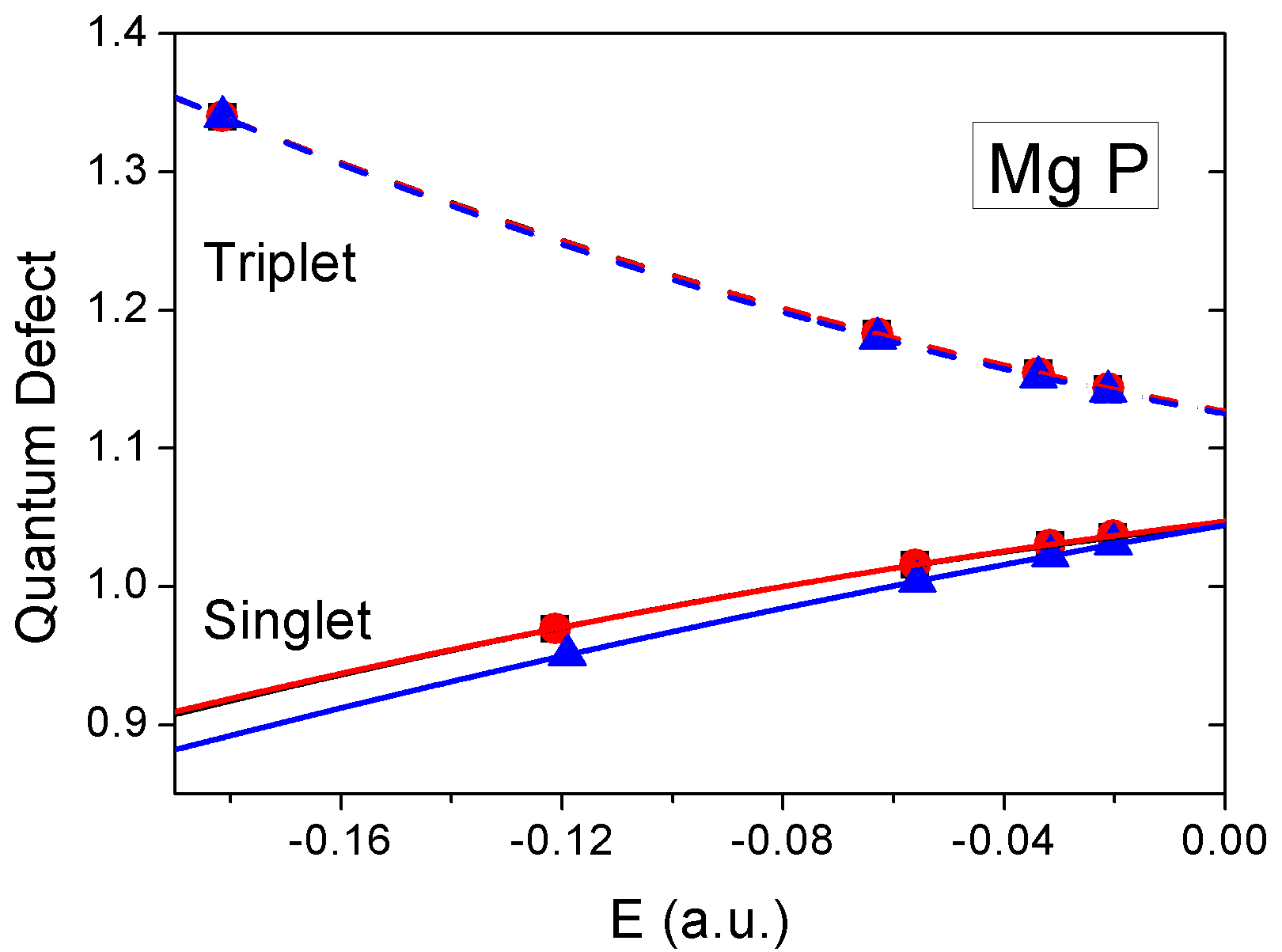}
\caption[]{\label{fig:MgP} Same as Fig. \ref{fig:BeP}, but for Mg.}
\end{figure}
\begin{figure}[ht]
 \centering
	  \includegraphics[width=0.5\textwidth]{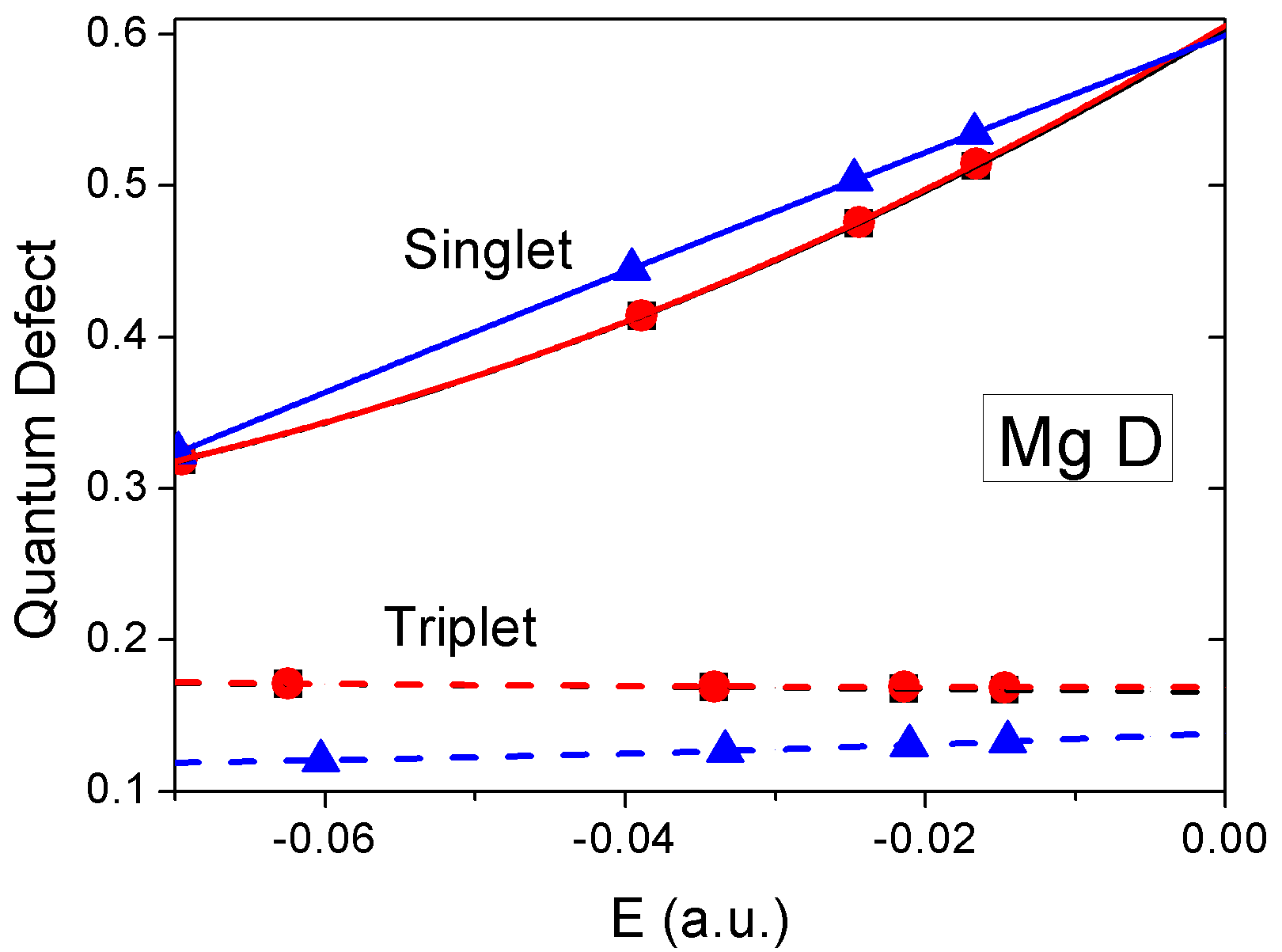}
\caption[]{\label{fig:MgD} Same as Fig. \ref{fig:BeD}, but for Mg.}
\end{figure}
\begin{table}
\caption[]{\label{QDMgApp} Same as Table \ref{QDexpt}, but for
pp Mg series.} 
\begin{ruledtabular}
\begin{tabular}{|c|ccc|}
tr. & $\mu_0$ & $\Delta\mu$ & $\Delta^2\mu$\tabularnewline
\hline
1S & 151.9  & -1.4  & 0.5 \tabularnewline
3S & 162.1  & -5.3  & 0.4 \tabularnewline
1P & 104.5  & 9.4  & -0.3 \tabularnewline
3P & 112.5  & -21.5  & 2.2 \tabularnewline
1D & 59.9  & 27.6  & -0.2 \tabularnewline
3D & 13.8  & 1.8  & 0.2 \tabularnewline
\hline
MSE & -0.6  & 0.8  & -0.3 \tabularnewline
MUE & 0.6  & 1.1  & 0.8 \tabularnewline
\end{tabular}
\end{ruledtabular}
\end{table}
To get an idea of how general this good performance of pp-RPA is, we
repeat the calculations for Mg.  In this case, the core is frozen
leading to a substantial underestimate of the IP
(0.13 eV) relative to experiment.  Using the experimental IP
with the pp transition frequencies would yield nonsensical
QD values.  But because our procedure fits this number, and then 
subsequent QD is plotted against the energy below threshold, we
can see the the Mg QD results are typically almost (not quite) as good as
those for Be.  
When averaging over all series, the Mg results are {\em better} than
those of Be.
This is a triumph of the QD method:  Despite a 
substantial error in IP, the QD is still extractable and the
underlying error in QD is very small.

\subsection{\label{Li} Results for an open shell}

We finish our survey with an extreme challenge for pp-RPA.
When the two-electron deficient reference is an unrestricted open-shell system, 
there are differences between $\alpha$ and $\beta$ orbitals. 
The spin contamination and potential spin incompleteness usually 
cause both pp-RPA and pp-TDA to produce meaningless results.
The Lithium atom is the simplest atoms with these problems.

With an $\alpha$-spin electron occupying the 1s orbital, the two-electron deficient reference for
Li is hydrogenic. In principle, we can create a
neutral doublet ground state by adding a $\beta$ electron to 1s
and another $\alpha$ electron to 2s, forming a $S_z=+\frac{1}{2}$ state.
But we could instead 
add two $\beta$ electrons to 1s and 2s separately to form the $S_z=-\frac{1}{2}$ state. 
Since we can only perform unrestricted calculations with our program, 
both series of excitations are spin-contaminated. 
We denote the two series as AB and BB. 

\begin{figure}[ht]
	  \includegraphics[width=0.5\textwidth]{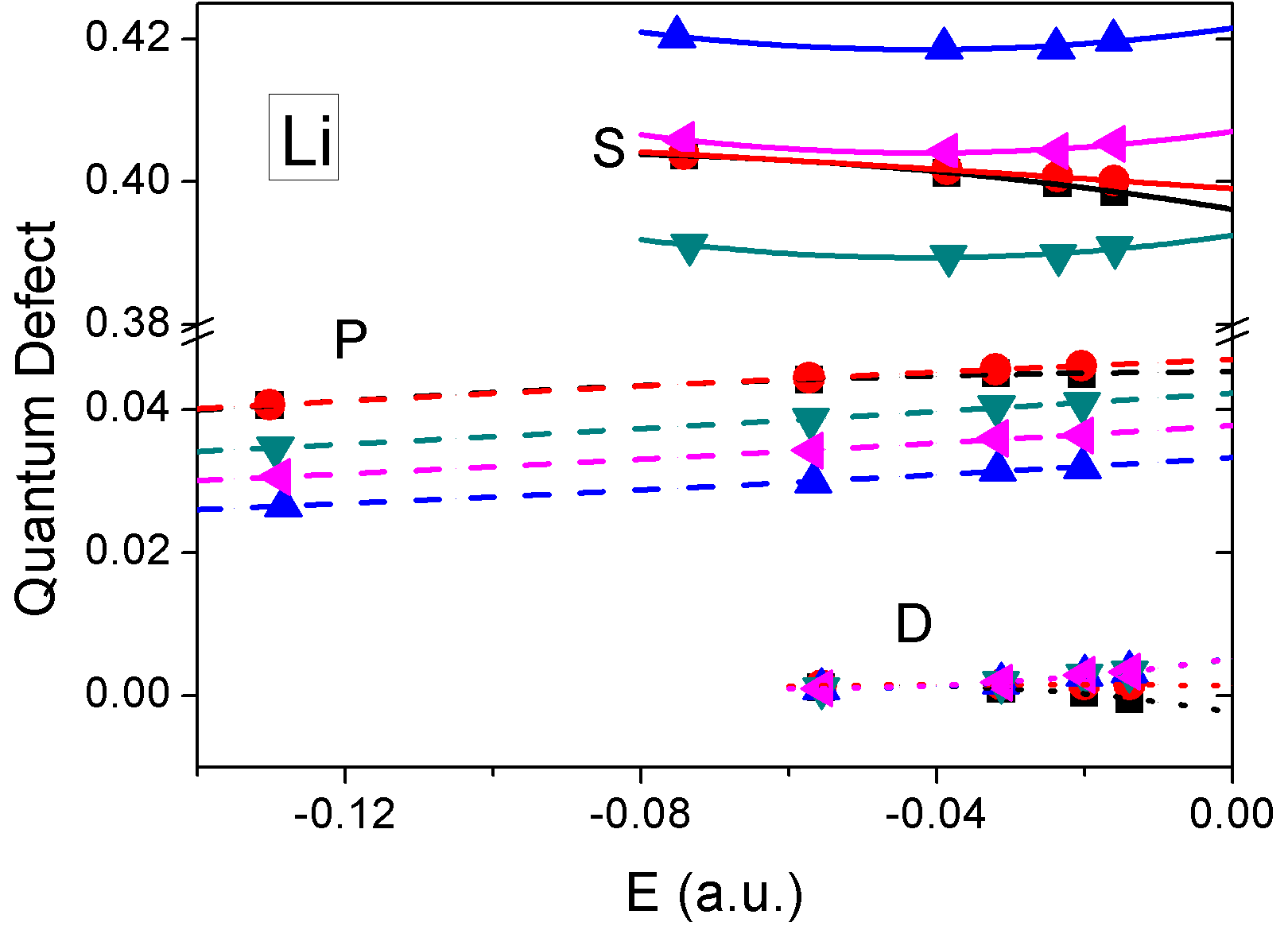}
\caption[]{\label{fig:Li} Quantum defects for atomic Li
from experiment (black), with fitted IP (red), from pp with AB (blue), with
BB (green), and averaging (pink).  Each series (S, P, D) has a distinct
line style.
}
\end{figure}
We can also attempt to overcome spin contamination by averaging these two, denoted AVE.
We try all three ways to compute the doublet S, P, and D series of Li. 
We saw in Section \ref{validity} that, since AB overestimates the IP and BB underestimates it,
their average has the most accurate IP, suggesting the AVE will yield
the best results for the transition energies too.

The results of all three series  are shown in figure \ref{fig:Li}. 
Note the $y$-scale is much smaller than in previous figures, and the
break in the curve.
Overall, all three sets yield fairly accurate QD results, with MUE's comparable to
those for Be and Mg.   For the S series, the AVE QD is substantially better
than either AB or BB, while for the D series, the three are indistinguishable.
But for the P series, the AVE curve is definitely worse than BB.  
So our QD analysis has shown that, although the AVE series always has the
most accurate transition energy (because of its accurate IP), it does
not always produce the best QD.

\begin{table}
\caption[]{\label{QDLiApp} Same as Table \ref{QDexpt}, but for
ppTDAHF Li series.} 
\begin{ruledtabular}
\begin{tabular}{|c|ccc|ccc|ccc|}
\multirow{2}{*}{tr.} & \multicolumn{3}{c|}{AB} & \multicolumn{3}{c|}{BB} & \multicolumn{3}{c|}{AVE}\tabularnewline
 \cline{2-4}\cline{5-7}\cline{8-10}
 & $\mu_0$ & $\Delta\mu$ & $\Delta^2\mu$ & $\mu_0$ & $\Delta\mu$ & $\Delta^2\mu$ & $\mu_0$ & $\Delta\mu$ & $\Delta^2\mu$\tabularnewline
 \hline
2S & 42.2  & 0.1  & 0.2  & 39.3  & 0.1  & 0.2  & 40.7  & 0.1  & 0.2 \tabularnewline
2P & 3.3  & 0.7  & 0.0  & 4.2  & 0.8  & 0.0  & 3.8  & 0.7  & 0.0 \tabularnewline
2D & 0.5  & 0.4  & 0.1  & 0.5  & 0.4  & 0.1  & 0.5  & 0.4  & 0.1 \tabularnewline
 \hline
MSE & 0.7  & 0.6  & 0.2  & 0.0  & 0.6  & 0.2  & 0.4  & 0.6  & 0.2 \tabularnewline
MUE & 1.5  & 0.6  & 0.2  & 0.5  & 0.6  & 0.2  & 0.9  & 0.6  & 0.2 \tabularnewline
\end{tabular}
\end{ruledtabular}
\end{table}

\subsection{\label{DFT}Importance of pp reference state}

The fact that pp-RPA and pp-TDA are able to describe charge
transfer and Rydberg excitations is often attributed to their
Coulomb and exchange kernels which 
are asymptotically correct. 
However, the orbital energies from
the two-electron deficient reference also play a vital role. 
Although the
Coulomb and exchange kernel is the same, the pp-RPA and pp-TDA with DFT references 
substantially underestimate the Rydberg excitation energies \cite{Yang_2013_224105} as a result of
the poor Kohn-Sham orbital energies, making it meaningless to
further look into the QD, just as in the TDDFT case. 
Because pp-RPA is only a first-order approximation to the 
adiabatic connection-pairing matrix fluctuation theory, it 
remains sensitive to the properties of the reference
calculation.  A well-behaved reference,
such as HF in this atomic Rydberg excitation case, is needed
for pp-RPA and pp-TDA to produce meaningful QDs.
However, for low-energy excitations in molecules, 
DFT references lead to significantly better results.\cite{Yang_2014_124104}

\section{Conclusion}

In this paper, we have developed a very general approach for extracting both threshold
energies and QDs from limited series of Rydberg excitations.   We have
demonstrated that our procedure can be used to extract extremely accurate
threshold energies (errors in the 0.1 mH range), and maximum errors of order 0.01
for QDs.  We have shown why measures like mean unsigned errors 
in collections of transition frequencies are not useful for Rydberg series,
and how to parametrize and quantify errors in QD's.   We find that pp-RPA
with a HF reference greatly outperforms TDDFT with a local approximation to the
exchange-correlation kernel.

The results reported here should become the benchmark for approximate calculations
of Rydberg series.   Our QD extraction procedure can be applied to other series,
such as excitions in solids.  Any quantum chemical method for excitations should
be measured against our pp-RPA results for these atoms.  We suspect they will
be difficult be beat with the same level of generality and low computational cost.

\begin{acknowledgments}
Y.Y. appreciates the support as part of the Center for the Computational Design of Functional Layered Materials, an Energy Frontier Research Center funded by the U.S. Department of Energy, Office of Science, Basic Energy Sciences under Award DE-SC0012575. K.B. was supported by 
the U.S Department of Energy (DOE), Of-
fice of Science, Basic Energy Sciences (BES) under award
DE-FG02-08ER46496. W.Y. was supported by the National Science Foundation (CHE-1362927).
\end{acknowledgments}

\end{document}